# Fixed-energy inverse scattering with radial basis function neural networks and its application to neutron–α interactions

Gábor Balassa [1]

[1] *Department of Physics, Korea University, 145 Anam-ro, Seongbuk-gu, Seoul 02841, Korea*
[2] *Institute for Particle and Nuclear Physics, Wigner Research Centre for Physics, Konkoly-Thege Miklós út 29-33, 1121 Budapest, Hungary*
*E-mail: balassa.gabor@wigner.hu

................................................................................

This paper proposes a data-driven method to solve the fixed-energy inverse scattering problem for radially symmetric potentials using radial basis function (RBF) neural networks in an open-loop control system. The method estimates the scattering potentials in the Fourier domain by training an appropriate number of RBF networks, while the control step is carried out in the coordinate space by using the measured phase shifts as control parameters. The system is trained by both finite and singular input potentials, and is capable of modeling a great variety of scattering events. The method is applied to neutron–α scattering at 10 MeV incident neutron energy, where the underlying central part of the potential is estimated by using the measured $l = 0$, 1, 2 phase shifts as inputs. The obtained potential is physically sensible and the recalculated phase shifts are within a few percent relative error.

................................................................................

Subject Index   A24, D00, D21

## 1. Introduction

Inverse scattering theory is a field that strongly couples mathematics and physics and has many applications in engineering [1,2], biology [3,4], and nuclear and particle physics [5–8]. In general, inverse scattering studies the characteristics and properties of an object by carrying out scattering experiments and measuring the changes in the excitations, which could be electromagnetic radiation or different kinds of particles, e.g. electrons, neutrons, protons, etc. The inverse scattering problem is often ill-conditioned and/or ill-defined; therefore, in many cases, it is a very hard task to solve it, even for relatively simple systems [9,10].

In this paper, low-energy elastic nuclear scattering is addressed, where an incoming particle is scattered on a target particle, without creating any other particles or being absorbed in the process. In low-energy inverse nuclear scattering, the interaction potential is usually sought by measuring different observables such as the total and differential cross sections, phase shifts, and polarizations [11]. Below the inelastic threshold, there will be no additional particles or inelastic effects, e.g. absorption, and the overall scattering problem is simplified; however, the inverse problem is still very hard to solve. There are different methods for grasping the inverse problem, each starting with different assumptions. One such technique uses the energy dependence of the observables to determine the scattering potentials at fixed angular momentum [12,13]; therefore, they are called fixed-angular momentum methods, e.g. Gelfand–Levitan–Marchenko inversion [14]. Another type of inversion uses the observables at different angular momenta and at fixed energies. These methods, such as Newton–Sabatier inversion, are called fixed-energy inversion [15,16]. There also exist so-called mixed inversion schemes, where energy dependence and angular momentum dependence are both used to estimate the scattering potentials [17]. Several extensions of the basic fixed-energy and fixed angular momentum methods also exist [18–20], where the mathematical formulation of the problem is at the forefront. On the other hand, with the fast growth and many modern developments of machine learning techniques [21,22], it would be interesting to see the possibilities of such techniques in a field that is quite far from the usual applications of "artificial intelligence." Here, we propose a data-driven inversion method using radial basis function (RBF) neural networks [23], where the scattering potential is estimated in the Fourier domain at fixed energy using the observed phase shifts of the different partial waves as inputs. The inversion potential is then corrected using an open-loop type control system, where, after inverse Fourier transformation of the potential into coordinate space, the measured phase shifts are used for correction with the help of the Simulated Annealing (SA) algorithm [24]. The method is based on two previous studies [25,26], where the inverse scattering problem is formulated as a system identification problem and noncausal Volterra series, and multilayer perceptron (MLP) neural networks were used to estimate the scattering potentials in low-energy nuclear scattering problems. In Ref. [26] the MLP construction is used to describe the $^3S_1$ NN potential from measured phase shifts within a few percent relative error accuracy. Here, a similar system is used; however, the problem is much harder because in the fixed-energy case, the possible inputs for the neural networks are limited, so a more careful approach is needed to be able to make reliable estimations.

The paper is organized as follows. In Sect. 2 the necessary mathematical preliminaries of the inverse scattering problem and the RBF networks are summarized, while in Sect. 3 the full RBF neural network model is described. After introducing the model, in Sect. 4 the network is trained with finite and singular potentials, where some examples are also shown. Finally, to use the model in a real-world scenario, in Sect. 5 it is used to describe neutron–$\alpha$ scattering at 10 MeV incident energies, where a physically sensible potential is achieved with a maximum relative error of a few percent in the recalculated phase shifts.

## 2.  Mathematical preliminaries

In this section, the necessary mathematical models are summarized, which will be used in the later parts of this paper. Starting with the problem of nuclear scattering experiments, we show that there is a nonlinear relationship between the observable phase shifts and the scattering potential, which ultimately leads to a nonlinear dynamical system representation of the inverse scattering problem. Systems that include nonlinear dynamics can be described by different methods, e.g. nonlinear Volterra series [27–29] or neural network models [30–32], for which we choose an RBF neural network representation. The main ideas and the mathematical preliminaries of the RBF networks are summarized here, while the specific model that will be applied to the inverse scattering problem will be described in Sect. 3.

### 2.1.  *Inverse scattering at fixed energy*

To describe the forward problem of nuclear scattering, one has to solve the 3D time-independent Schrödinger equation, which is a linear, second-order partial differential equation that describes the evolution of the $\Psi(\mathbf{r})$ wave functions under a scattering potential $V(\mathbf{r})$ [33]. Assuming spherical symmetry and factorizing the wave functions into radial and angular parts, the problem is simplified into solving the radial Schrödinger equation, defined as:

$$-\frac{\hbar^2}{2\mu}\frac{\partial^2 u_l(r)}{\partial r^2} + \left[V(r) + \frac{\hbar^2}{2\mu}\frac{l(l+1)}{r^2}\right]u_l(r) = Eu_l(r) \tag{1}$$

where $\mu = m_1 m_2/(m_1 + m_2)$ is the reduced mass of the interacting particles, $V(r)$ is the radially symmetric potential, $\hbar$ is the reduced Planck constant, $E$ is the energy, while $l$ is the angular momentum quantum number. The radial wave function in this case is defined as $u_l(r) = rR_l(r)$, where $R_l(r)$ is defined by the factorization of the wave function $\Psi(\mathbf{r}) = R_l(r)Y_l^m(\theta, \phi)$.

In an elastic scattering scenario, the incoming wave function, which represents the bombarding particle, will be perturbed by the target particles. In scattering experiments, only the asymptotic behavior can be measured; therefore, we are also interested in the asymptotic behavior of the wave functions. The general procedure is to assume that the asymptotic wave function is the sum of an incoming plane wave and an outgoing spherical wave, which is weighted by the so-called scattering amplitude $f(k, \theta)$ as:

$$\Psi(\mathbf{r})_{AS} = e^{i\mathbf{k}\mathbf{r}} + f(k, \theta)\frac{e^{ikr}}{r}. \tag{2}$$

By applying a partial wave expansion and with some manipulation, the scattering amplitude can be expressed in the following form [34]:

$$f(k, \theta) = \frac{1}{2ik}\sum_{l=0}^{\infty}(2l+1)(e^{2i\delta_l}-1)P_l(\cos(\theta)), \tag{3}$$

where $P_l(\cos(\theta))$ is the $l$th order Legendre polynomial, and $\delta_l$ is the phase shift of the $l$th partial wave. In scattering experiments, the differential and total cross sections are measured, where the differential cross section can be defined from the scattering amplitude as:

$$\frac{d\sigma}{d\theta} = |f(k, \theta)|^2. \tag{4}$$

By measuring the differential cross sections, the phase shifts for the respective partial waves can be deduced by a fitting procedure. The relationship between the phase shifts and the potential can be further examined by the variable phase approximation (VPA) [35–37], which describes the relation between $\delta_l$ and $V(r)$ with a nonlinear first-order differential equation derived for the so-called phase function of the $l$th partial wave:

$$\frac{d\delta_l(r)}{dr} = -\frac{2\mu V(r)}{k\hbar^2}\Big[j_l(kr)\cos(\delta_l(r)) - n_l(kr)\sin(\delta_l(r))\Big]^2, \tag{5}$$

where $\delta_l(r)$ is the phase function at the distance $r$, $k$ is the center-of-mass momentum defined as $k = (2\mu E/\hbar^2)^{1/2}$, while $j_l(kr)$ and $n_l(kr)$ are the Ricatti-Bessel and Ricatti-Neumann functions, respectively. The center-of-mass momentum can be expressed by the laboratory kinetic energy $T_{\mathrm{lab}}$ as follows:

$$k = \frac{m_2^2\left(T_{\mathrm{lab}}^2 + 2m_1 T_{\mathrm{lab}}\right)}{(m_1 + m_2)^2 + 2m_2 T_{\mathrm{lab}}}, \tag{6}$$

where $m_1$ is the mass of the incident particle, while $m_2$ is the mass of the target particle. In real scattering experiments, only the asymptotic phase shifts are accessible; therefore, in Eq. (5) the $\delta(r \to \infty)$ value is of interest. In practice, the VPA equation is relatively hard to solve, especially for higher angular momenta, and very precise methods, e.g. 5th-order Runge-Kutta, are necessary. Considering the forward problem of the experimental setup, it is a highly nonlinear, dynamical system, in which case the inverse problem will also be a very difficult nonlinear system to be solved. The analytical formulation of such a problem is very hard; however, as we will see later, with the help of techniques used mostly in engineering applications, it is possible to estimate the solution for the inverse problem with good accuracy in a well-defined operating range, which in this case will mean possible constraints to the potentials, e.g. maximum and minimum values, range, maximum "frequency," or intervals of the possible phase shifts. As the proposed model is a so-called gray box, where preliminary information is used in the identification process, we have great flexibility on how to approach the problem. In the next section, we briefly introduce the RBF-type neural network structure, which will be used to solve the inverse scattering problem.

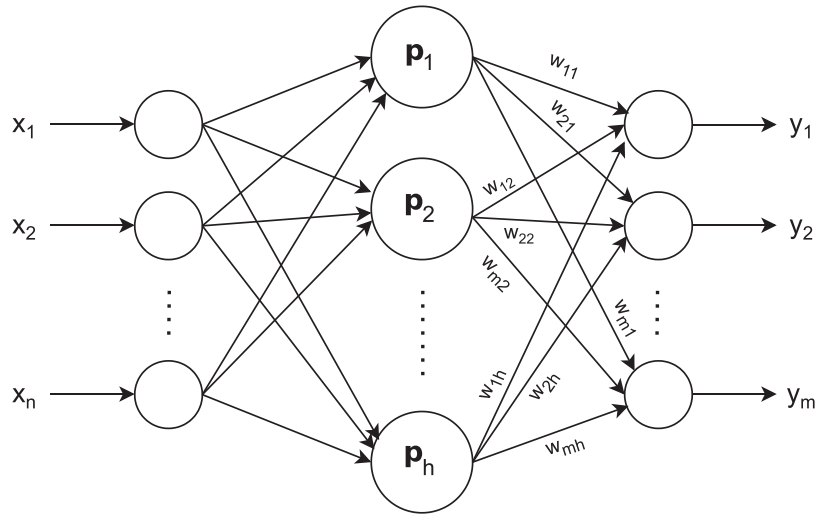

**Fig. 1.** Schematic representation of an RBF neural network.

### 2.2. *RBF neural networks*

Machine learning techniques and neural networks are one of today's foremost research topics in modeling complex systems, where the underlying mathematics/physics are not well understood [38–40] or when the underlying system is fairly well understood but a fast, approximate solution is more desirable [41–43], e.g. in aerospace engineering [44], where a good neural model could speed up the modeling step by hundreds of times.

There are numerous neural network models available, e.g. MLPs [45], support vector machines (SVMs) [46], invertible neural networks (INNs) [47], convolutional neural networks (CNNs) [48], etc., each with their own advantages and drawbacks. Determining which model is best for a given application is an essential step in the modeling process. One of the earliest neural network models in the field is called the RBF neural network, which has a feed-forward type network structure with one input-, one output-, and one hidden layer. The hidden layer is made up of nonlinear radially symmetric activation functions, while the output layer implements a weighted linear relationship between the output of the hidden layer and the output of the full system. The schematic structure of an RBF network can be seen in Fig. 1, which shows a multiple input-multiple output system (MIMO). Following the notations shown in Fig. 1, the mathematical representation of an RBF network can be summarized as follows:

$$y_k(\mathbf{p}_1, \mathbf{p}_2, ..., \mathbf{p}_h; \mathbf{x}) = \sum_{i=1}^{h} w_{ki} F_i(\mathbf{p}_i; \mathbf{x}), \qquad (7)$$

where $\mathbf{x} = (x_1, x_2, ..., x_n)$ are the inputs of the system, $y_k$ is the $k$th output ($k = 1...m$), while $F_i(\mathbf{p}_i; \mathbf{x})$ represents the RBFs, which depend on the inputs and on the $\mathbf{p}_i$ parameters. In the special case when the RBFs are all Gaussians, the parameters $\mathbf{p}_i = (\mathbf{c}_i, \sigma_i)$ will be the centers $\mathbf{c}_i = (c_{i_1}, c_{i_2}, ..., c_{i_n})$, and the widths $\sigma_i$ and the basis functions will have the following form:

$$F_i(\mathbf{p}_i; \mathbf{x}) = F_i(\mathbf{c}_i, \sigma_i; \mathbf{x}) = e^{-\frac{||\mathbf{x}-\mathbf{c}_i||^2}{2\sigma_i^2}}, \tag{8}$$

where in this case we will take the 2-norm when calculating the $|| \cdot ||$ distances. Here, the $F_i(\mathbf{c}_i, \sigma_i; \mathbf{x})$ function represents one RBF, with the parameters $\mathbf{c}_i$ and $\sigma_i$ taking the full $\mathbf{x} = (x_1, x_2, ..., x_n)$ values as inputs, while each $y_1, y_2,..., y_m$ output will be a linear combination of the $F_1(\mathbf{c}_1, \sigma_1; \mathbf{x})$, $F_2(\mathbf{c}_2, \sigma_2; \mathbf{x}),..., F_h(\mathbf{c}_h, \sigma_h; \mathbf{x})$ RBFs. It is also possible to use different basis functions and/or different norm definitions, which could be advantageous in some specific situations. RBF networks have been intensively studied in the past and have numerous applications in system identification, control systems, chatotic time series prediction, speech and image recognition, and many more [49–52]. In the mathematical sense, it can be proven that the RBF networks are universal approximators [53], therefore they can be used to model a wide range of complex mappings between the inputs and the outputs. The free parameters of the model are the $\mathbf{c}_i$ centers, the $\sigma_i$ widths, and the $w_{ik}$ linear weighting factors, which have to be determined by a chosen training method. Here, one has many possibilities. The most straightforward way to train the network is to train all the parameters in a supervised fashion and use, e.g. the backpropagation algorithm to find some local minimum in the cost function [54]. In this case, one has to find optimal centers, weights, and linear weights by iteratively searching for better solutions. While this method gives a more flexible construction, it is also numerically harder to achieve a good, fast, and robust estimation for the parameters. Without good regularization, the parameters could cover a very wide range, which in turn could lead to a very sensitive approximation with bad generalization capabilities. The time for a good backpropagation learning could also be very long, depending on the number of basis functions, as each kernel would have three trainable parameters and the overall output would depend nonlinearly on all the widths and centers. Another way to train the RBF networks is to use unsupervised learning at least for some of the parameters, typically for the $c_i$ centers and for the $\sigma_i$ widths [55,56]. In this case, the parameters on which the output depends nonlinearly are trained in an unsupervised manner by some clustering or classification algorithm, while the linear weights are trained in a supervised manner, e.g. by the backpropagation algorithm or by simple matrix inversion. Due to the fact that output depends linearly on the trainable weights, the overall optimization problem is better defined. In this case, the training process could be significantly shorter; therefore, it is a good way if one wants to use RBF networks in real-time applications. One thing to keep in mind is that in each case, the generalization capabilities have to be addressed, and a careful examination is needed, e.g. in the case of unsupervised learning, one has to be sure that the parameter hypersurface is sufficiently covered so that the network can give a good, continuous response to the excitations.

## 3.   Neural network model for fixed-energy inversion

To solve the inverse problem of low-energy elastic particle scattering, an RBF-based neural network model is constructed, which uses the measured phase shifts as inputs to determine the interaction potential in the Fourier domain. As we will only consider radially symmetric

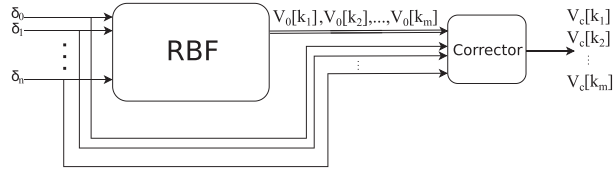

**Fig. 2.** Schematic view of the inversion method. On the first step the RBF neural network estimates the $V_0(k)$ potential, then in the next step the corrector will fine-tune the potential to the measured phase shifts.

potentials, the 3D Fourier transform f($k$) of a radial function f($r$) can be written as:

$$f(k) = \mathcal{F}\Big[f(r)\Big] = \frac{4\pi}{k} \int_0^\infty r \sin(kr) f(r) dr. \tag{9}$$

Following the definition of the Fourier transform, the inverse transform can be defined as:

$$f(r) = \mathcal{F}^{-1}\Big[f(k)\Big] = \frac{1}{2\pi^2 r} \int_0^\infty k \sin(kr) f(k) dr. \tag{10}$$

In practical calculations, we only consider potentials that decay fast enough so that the above integrals can be done without any difficulty; however, it has to be noted that in general, the integral could be very hard to evaluate due to the strongly oscillating behavior at large $k$'s and large $r$'s. It is also worth noting that the usage of the radial Fourier transform is only a modeling choice and it has no physical significance, as we could easily choose, e.g. the 1D Fourier transform for the radial distance (symmetrized at $r = 0$ so that the integral transform becomes purely real) instead. It is also a modeling step to determine how many partial waves we need to be able to achieve good accuracy. In theory, it is necessary to use all of the partial waves to be able to precisely determine the scattering potentials; however, in practice, we are dealing with a constrained version of the full problem as we will imply specific constraints on the potentials. Keeping this in mind, the actual number of phase shifts we need will be determined by training and comparing different networks with different numbers of input phase shifts.

The schematic model of the inversion method can be followed in Fig. 2. The first step of the inversion is labeled by "RBF" and its purpose is to give a preliminary estimate of the potential $V(k)$, which is the radial Fourier transform of the $V(r)$ interaction potential. The RBF network will be trained by using generated training data, and it is expected that the $V_0(k)$ estimate will be close to the true $V(k)$; however, as the inverse problem is very sensitive even to small perturbations and the inverse Fourier transform could also introduce small perturbations into the estimated $V_0(r)$ coordinate-space potential, we also include a corrector step just after the RBF network, whose purpose is to correct the $V_0(r)$ estimate so that the recalculated phase shifts are close to the measured phase shifts. This corrector step closely resembles a so-called open-loop control, where the output of the system is controlled without a feedback loop by only using information that is accessible beforehand. In this case, it is quite natural to use the measured phase shifts as, in the end, those are the observables we want to describe with the inversion technique.

Let us first describe the RBF network, which will be used to give an estimation of the interaction potential in Fourier space. The schematic of the method can be seen in Fig. 3, where the input phase shifts are marked by $\delta_i$, and the potentials are evaluated at a disretized $k_i$ grid. Each discrete $V(k_i)$ will have its own RBF subnetwork (RBF$_i$), and the output of each RBF subnetwork will be connected to the input of the subsequent RBF network, which will be used to estimate the potential at the next $k_{i+1}$ grid point. A good estimation of the potential in the first point $V(k_1)$ is critical as it only has a finite number of phase shifts as inputs and no other information will be available to it. The subsequent points will have extra knowledge of the shape of the potential as well, which will greatly help in the training process. The block diagram in Fig. 3 shows a specific case where each RBF subnetwork will use all the previous outputs of the subnetworks as input; however, as we will see in Sect. 4, it is not necessary to use all the available previous outputs, and a finite memory case will be enough. To be able to fully make use of the information stored in the shape of the potentials, it is necessary to use a fine grid, e.g. a sufficiently small $\Delta k$. It is worth mentioning that the full grid does not have to be uniformly distributed, and we could make use of the preliminary information given to us by the knowledge of the training potentials, which will be further elaborated in Sect. 4. In the mathematical sense, the relationship between the inputs and outputs can be written as follows:

$$V(k_1) = \text{RBF}_1(\delta_0, \cdots, \delta_n) \tag{11}$$

$$\vdots \tag{12}$$

$$V(k_m) = \text{RBF}_m(\delta_0, \cdots, \delta_n, V(k_{m-1}), \cdots, V(k_{m-M})), \tag{13}$$

where the $\text{RBF}_i$ represent the functional form of the $\text{RBF}_i$ subnetwork, while $M$ is the memory of the system. According to Eq. (8), each $\text{RBF}_j$ subnetwork can be given by the following form:

$$\text{RBF}_j(\mathbf{x}) = \sum_{i=1}^{h_j} w_{ji} e^{-\frac{\|\mathbf{x} - \mathbf{c}_{ji}\|^2}{2\sigma_{ji}^2}}, \tag{14}$$

where $\mathbf{c}_{ji}$ are the centers, $\sigma_{ji}$ are the widths, $w_{ji}$ are the weights, while $h_j$ is the number of Gaussian kernels in each subnetwork, and $\mathbf{x} = (\delta_0, \ldots, \delta_n, V(k_{m-1}), \ldots, V(k_{m-M}))$ are the inputs of the actual subnetwork, where, e.g. $V(k_3) = \text{RBF}_3(\ldots)$. The $j$ indices in $c_{ji}, \sigma_{ji}$, and $w_{ji}$ mean that in every subnetwork we will have the freedom to choose different centers, widths, and weights, which is necessary as each following subnetwork will take the outputs of some of the previous subnetworks as inputs (defined by the $M$ memory), thus changing the input space that has to be covered.

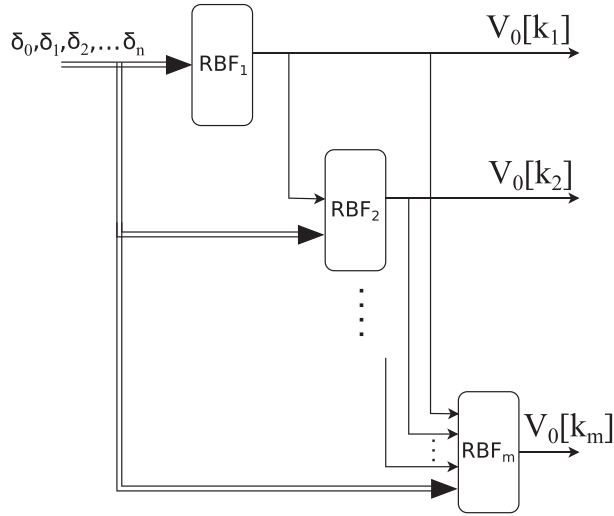

**Fig. 3.** Block diagram of the RBF inversion (the detailed model of the "RBF" block in Fig. 2) for the full-memory case, where each subnetwork will use the outputs of all the previous subnetworks.

In general, one has great flexibility in choosing the grid and the memories. As stated before, the grid does not necessarily have to be uniformly distributed, but we could choose a finer grid where we expect faster changes and a sparser grid where we expect a slowly varying form. The memory also does not have to be the same in all of the grid points; however, for simplicity and consistency, in the following, we will always choose a constant $M$.

The RBF$_i$ nonlinear activation functions are given by Gaussians with the form shown in Eq. (14) with the $\mathbf{c}_{ji}$ centers and $\sigma_{ji}$ widths as free parameters, which have to be determined. There are many methods for determining these parameters. In this work, we use nonsupervised learning to determine the centers and widths and supervised learning to determine the linear weights in Eq. (7). To determine the centers, the k-means clustering algorithm [57] is used, which is a simple yet effective algorithm that separates the data into $k$ clusters, where the cluster means represent the centers that we want to determine. In practice, it is best to check the model accuracy with a varying number of centers and make a decision by taking into consideration the obtained errors and the complexity of the whole system. Here, we will do the same and train the model with different numbers of centers to be able to decide which is the optimal model, which is not too complex (training time is not too large, no overfitting, etc.), but not too simple (error is not too large), and could give a good overall performance.

After the centers are fixed, one has to choose appropriate weights as well. This is also not a trivial task, as in the case of too small weights, we risk that the input hyperspace will not be covered or the model will overfit to the nearest points, while if the weights are too large, the model will not be able to learn the subtle changes of the system. The weights also do not have to be constant, and each weight could be different, in which case the best solution would be to train the weights in a supervised manner. In practice, one wants a sufficient but not too large overlap between the Gaussian basis functions so that smooth interpolation can be achieved. During the training of the proposed model, many possible weight configurations were studied, and in many cases, the best performances were achieved when the weights were determined by averaging the distance between the nearby centers as follows:

$$\sigma_i = \frac{1}{2(N-1)} \sum_{j=1 (j \neq i)}^{N} \sqrt{||\mathbf{x}_i - \mathbf{x}_j||^2}, \tag{15}$$

where $N$ is the number of centers, and $\mathbf{x}_i$ is the $i$th input sample, which in the case of the first RBF subnetwork will consist only of phase shifts, e.g. $\mathbf{x}_i = (\delta_{0,i}, \delta_{1,i}, \ldots, \delta_{n,i})$. An example of the calculated widths for four 2D centers can be seen in Fig. 4. It can be seen that with this method, the overlap is not too large; however, there is still a significant overlap between the points, which is exactly what we want from the weights.

To be able to find the centers and widths accurately, it is necessary to normalize (or standardize) the data. In this case, all the inputs (and outputs) are normalized between $[-1,1]$ and the errors that will be shown in the following sections are all calculated by the normalized values. After the centers and the widths are determined, the linear weights are trained in a supervised manner by using generated training data. The training procedure and the generation of the training samples will be the main topics of Sect. 4, where the model complexity will be further analyzed. This concludes the first part of the inversion shown in Fig. 3, and now we can pro-ceed further to the description of the corrector step, which aims to fine-tune the obtained $V_0(k)$ potentials.

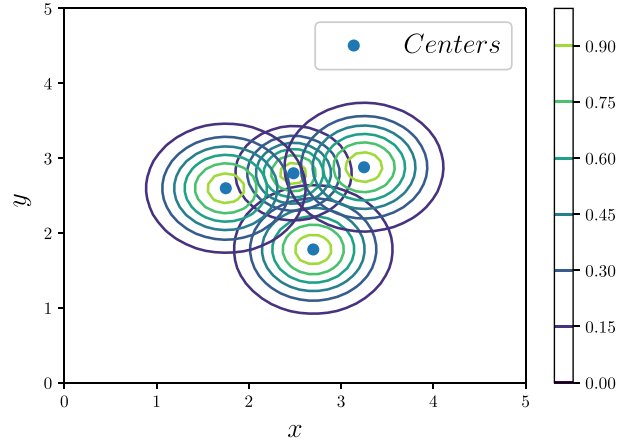

**Fig. 4.** Overlap of four RBFs with the calculated $\sigma_i$ parameters from Eq. (15).

It was mentioned before that the calculated phase shifts are very sensitive to the small perturbations of the potentials, and even if the overall shape of the potential is good, it is possible that the recalculated phase shifts have large errors. To avoid this, a simple corrector step is included in the inversion procedure, which, by making small changes in the inverse Fourier transformed $V_0(r)$ potentials, fits them to the measured phase shifts. This method will be quite flexible, as we do not necessarily have to use the same number of phase shifts as that which is used for the RBF network as inputs, but we could use all the available measured phase shifts as well. The corrector block therefore will take the estimated $V_0(r)$ inverse Fourier transformed potential in the coordinate space and some of the measured phase shifts $\delta_i$ as inputs, and by making small perturbations in $V(r)$ it will correct it step-by-step to make sure that the final recalculated phase shifts $\delta_{c,i}$ are close to the measured ones. There are many ways one could make small changes to the potential; however, we would want to make sure that the resultant potential satisfies some constraints, e.g. continuity, boundedness, etc. Here, we will use a Gaussian RBF description of the potential so that it is consistent with the preceding neural network model. In this picture, we first expand the coordinate-space potential with the help of Gaussian basis functions, similar to how one does it with the RBF network. Now, however, we have more control and more information about the functional form we want to represent; therefore, in this case, we will use the neural network structure for a simple interpolation as follows:

$$V(r) \approx \sum_{i=1}^{N} a_i e^{-\frac{(r-c_i)}{2\sigma_i^2}},\qquad(16)$$

where $a_i$, $c_i$, and $\sigma_i$ are the usual free parameters of the Gaussian representation. The potentials we would like to describe have a dominant contribution below $r \approx 5$ fm, so we could choose the centers so that they are distributed in this region. In practice, one has to choose the centers and widths so that the potential can be interpolated smoothly, but it is usually not a problem because we train our network for "well-behaved" potentials. In fact, it is perfectly enough to distribute, e.g. $N = 30$ centers evenly between 0 and 5 fm and choose the widths so that $\sigma_i = 1$ fm for all $i$. After we "train" this network (the parameters $a_i$ are fitted) we want to make small changes in them, so that the resultant $V(r)$ will also be changed. More precisely, we will add small $\Delta a_i$ perturbations to the $a_i$ parameters so that we get a smooth $\Delta V(r)$ perturbation in $V(r)$ as follows:

$$\Delta V(r) = \sum_{i=1}^{N} \Delta a_i e^{-\frac{(r-c_i)}{2\sigma_i^2}} , \tag{17}$$

where one has to make sure that the perturbations in $a_i$ are sufficiently small so that the resulting perturbations in $V(r)$ will be well-constrained as well. The aim of changing the parameters in the Gaussian representation is to find a (local) minimum of a cost function corresponding to the difference between the measured and the recalculated phase shifts. Here, the cost function is defined as the averaged relative difference of the phase shifts, defined as:

$$\hat{E}_\delta = \frac{1}{L_{\max}} \sum_{l=0}^{L_{\max}} \frac{|\delta_l - \delta_{c,l}|}{|\delta_l|}, \tag{18}$$

where $L_{\max}$ is the number of phase shifts considered, $\delta_l$ are the measured phase shifts, and $\delta_{c,l}$ are the phase shifts calculated from the estimated potential. To reach an optimum in the cost function, the SA algorithm is used [58], which is well-suited for problems where, e.g. gradient methods are not feasible. In the SA algorithm, one starts with an initial guess for the optimizable parameters, and then by making random changes to them, the new parameters will be accepted if a better solution is achieved; however, these parameters will also be accepted with some predefined probability (acceptance probability) if the new results are worse than the previous ones. This probability can be defined in many ways, and usually it is defined so that with each step the acceptance probability will be smaller, so the probability of accepting a wrong answer will be smaller with each step. This formalism, however, will help the algorithm avoid getting stuck in a local minimum. The actual form of the acceptance probability could be defined in many ways, with different speeds of decrease and different functional forms. Here, we apply an exponentially decaying acceptance probability defined as:

$$T = T_0 e^{-kn}, \tag{19}$$

where $T_0$ and $k$ are free parameters, and $n$ is the number of iterations. The free parameters control the speed of convergence, and their values usually depend on the problem at hand. In the following problems, their values are set to be $T_0 = 1$ and $k = 0.15$; however, it is also possible to apply an adaptive algorithm to decide them during the control step. It is also possible to choose a different functional form, e.g. linearly decreasing, which could give better convergence for some problems as well. The exponentially decreasing functional form makes sure that the acceptance probability of worse configurations will be smaller with each step; therefore, at sufficiently far from the starting point, the optimizer will (almost) only accept better parameter configurations. For the stopping condition, we will check the averaged relative error at each step, and if it is smaller than 0.01, or the number of iterations is larger than $10^4$, the optimization will be stopped. As the SA optimizer is very flexible, it is possible to fit the stopping conditions to the actual problem, e.g. if the errors of the measured phase shifts are larger than a few percent, then it does not make sense to try to give a 1% accuracy with the inversion. Therefore, the actual value should depend on the measurement errors of the phase shifts we will use in the control step.

In the next section, the inversion procedure will be put to use by training the RBF network to radially symmetric local potentials and applying the inversion and the SA optimizer to different test potentials.

## 4. Inversion for finite and singular potentials

To train the RBF network, it is necessary to construct well-defined training potentials. As the forward problem is well-defined, it is relatively easy to calculate the phase shifts for radially symmetric local potentials by solving the VPA equations described in Eq. (5). As we would like to describe nuclear potentials, which could have an infinitely large repulsive part near the origin due to the charged Coulomb force and Fermi exclusion, it is necessary to include these types as well in the training procedure. In Sect. 4.1, we will only use finite potentials; in Sect. 4.2, we extend the training samples by potentials which could have large repulsive parts near the origin.

### 4.1. *Finite potential case*

In system identification, it is an important task to design excitation signals that will excite the system within a well-defined operating range with specific features. In practice, it is also desirable to maintain a certain randomness of the signals to prevent overtraining with a specific type of excitation. In frequency domain analysis, one would like to excite all the possible frequencies related to the bandwidth of the system in question, for which random noise signals, swept sine signals, impulse excitation, or different kinds of multisignals, etc. could be used as well [59–62]. In linear systems, the problem is drastically simplified as there is no frequency doubling or other nonlinear effects, which would otherwise make the identification process much more difficult. In our case, we would like to generate random potentials that have specific features such as boundedness, continuity, finiteness, etc. The general properties of the expected potentials could be estimated by analyzing the underlying problem we would like to solve, which in this case means low-energy, elastic nuclear scattering. In this case, if we omit the possible large repulsive part at the center, we expect potentials that die out around 4–6 fm and have finite values between a few hundred MeV; therefore, these will be our main guides when we generate the training samples for identifying the inverse system. The training samples are generated by so-called random phase multisines, which have many good properties when the underlying system has nonlinearities that have to be detected during the system identification process [63,64]. Random phase multisines are sums of harmonic functions with varying phases and amplitudes, usually generated in the frequency domain by defining the amplitude spectrum first. The functional form can be expressed as:

$$V_{ms}(r) = \sum_{i=1}^{N} a_i \sin(k_i r + \phi_i), \tag{20}$$

where $N$ is the maximum number of harmonics, $k_i$ is the "frequency," while $a_i$ is the amplitude of the $i$th harmonic, and $\phi_i$ is the random phase, chosen randomly from a uniform distribution between 0 and $2\pi$. The frequency components could be chosen in many ways, e.g. only odd or only even frequencies, and a specific configuration could greatly help the identification process, especially when one wants to identify nonlinearities in the system. Here, we will use a uniformly distributed frequency grid with $k_i = i\Delta k$. The multisine signal could excite the system at a range of $k_i$'s; however, we also expect that the potentials will die out around $r \approx 4$–6 fm, so the multisine excitations will be multiplied by a decaying Gaussian factor to make sure that at large distances the potentials will be negligible:

$$V_0(r) = A_N \frac{V_{ms}(r)}{\max[|V_{ms}(r)|]} e^{-cr^2}, \tag{21}$$

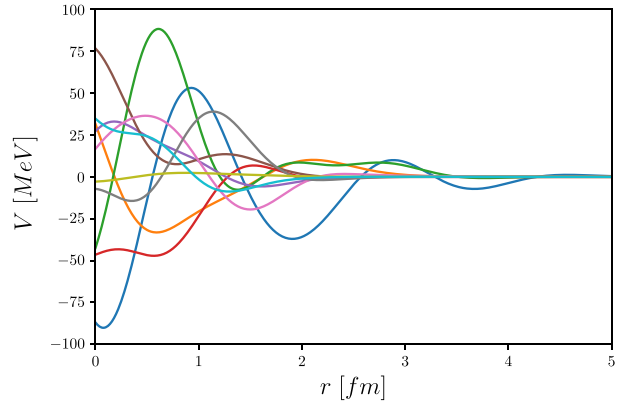

**Fig. 5.** Decaying multisine training potentials defined in Eq. (21).

where the random phase multisine is normalized so that it always stays between $[-A_N, A_N]$, while the $c$ parameter of the decaying factor is set so the potentials will die out around a few fm. To generate the training potentials, the parameters have been set to $k \in [0, 5]\,\text{fm}^{-1}$, $\Delta k = 0.1$ fm$^{-1}$, $N = 50$, $a_i \in U[0, 1]$, $\phi_i \in U[0, 2\pi]$, $c \in U[0.2, 1]$, $A_N \in U[-100, 100]$ MeV. In practice, first we choose a random number between 1 and $N = 50$, which will define the number of harmonics, then we choose $N$ number of $\phi_i$ random phases, $a_i$ amplitudes, the $A_N$ normalization factor, and the $c$ decaying factor uniformly from the predefined ranges. Multiplying the multisine signal by a Gaussian decaying factor in coordinate space means a convolution in Fourier space; therefore, we have to consider $k > 5$ fm$^{-1}$ values in the identification process as well. In Fig. 5 some training potentials can be seen that were generated by the method described here.

In this section, we will set the system so that it resembles simple nucleon–nucleon scattering by setting the masses of the colliding particles to $m_{1,2} = 940$ MeV, while the laboratory energy is set to be $T_{\text{lab}} = 10$ MeV.

To train the RBF network, 10,000 potentials were generated, while for validation, 2000 samples were used. To compare the performance of the networks with different complexities, the root mean squared errors (RMSEs) were calculated at each inversion point, then averaged to be able to give one number that measures the performance of the different systems; thus, the error measure that will be used from now on is defined as:

$$\text{RMSE} = \frac{1}{N_I} \sum_{i=1}^{N_I} \left( \sqrt{\frac{1}{N_V} \sum_{j=1}^{N_V} \left( \tilde{V}_N(k_j) - \tilde{V}_N^0(k_j) \right)^2} \right), \tag{22}$$

where $N_I$ is the number of inversion points, $N_V$ is the number of samples used for validation, $\tilde{V}_N(k_j)$ is the normalized model calculation at $k_j$, while $\tilde{V}_N^0(k_j)$ is the normalized true value of the potential. The inputs and outputs are normalized to $[-1,1]$ so it is not necessary to use a normalized error measure.

The two main parameters of complexity are the number of input phase shifts and the number of Gaussian basis functions (number of centers), for which many networks have been trained. The first inversion point at $k = 0$ fm$^{-1}$ is crucial as all the follow-up points will use it as an input, so at first, the phase shift and number of center dependence have been examined only for the first inversion point, which can be seen in Fig. 6. In this case, $N_I = 1$, as only the first inversion point is evaluated. In Fig. 6 each color corresponds to a different number of input phase shifts, e.g. the blue line only uses the s-wave, while the black line uses the s, p, d, f, and g wave phase shifts as well. It can be seen that with each additional phase shift, the performance will be better; however, after using the s, p, and d phase shifts, the RMSE does not decrease a lot. This tendency can also be seen in the RMSE dependence on the number of centers $N_c$, where after $N_c = 1000$ the RMSE will decrease very slowly, and after a few thousand centers we reach a point where the network is overtrained and the validation error will get larger or stagnate. This point is different for networks with different numbers of phase shifts, e.g. when the s, p, d, and f phase shifts are used, we still get a slowly decreasing tendency at 3000 centers. It is worth noting that the more training samples we have, the more centers we could use without overtraining; however, for practical purposes, in our case, 10,000 samples were enough to show the capabilities of the proposed model. In Fig. 7 a comparison of the normalized data is shown for a network with 3 input phase shifts (s,p,d) and 3000 centers, where the red dots represent the model calculations and the black lines are the data used for validation. These results correspond to RMSE $\approx 4 \cdot 10^{-3}$ and while there is some discrepancy at some points, the overall performance is good enough. These are important results, as in many real measurements, we only have access to two or three measured phase shifts, so it is desirable to use the least amount of phase shift possible which could give us a satisfying performance. In this case, we will use only the s, p, and d phase shifts as inputs; however, it has to be mentioned that if we have a more complex problem with, e.g. potentials with a larger effective distance, we would need more phase shifts, and this has to be examined carefully in each problem.

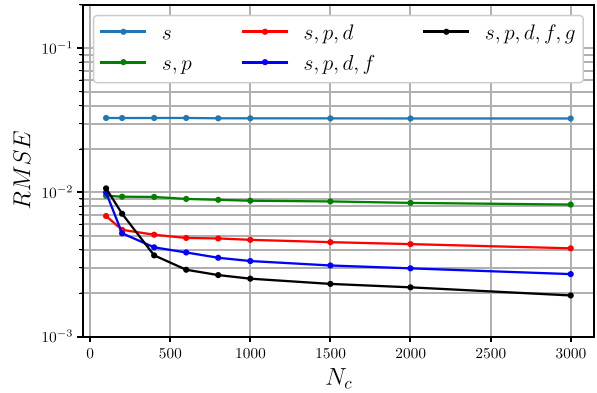

**Fig. 6.** RMSE dependence of the normalized $\widetilde{V}_N(k = 0)$ on the number of basis functions, and input phase shifts for finite training potentials.

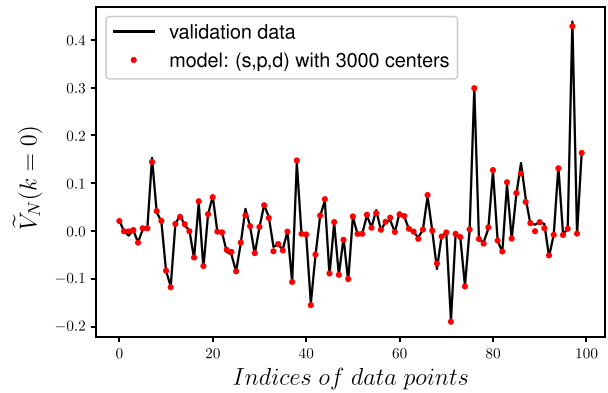

**Fig. 7.** Comparison of the true and estimated (normalized) potentials at $k = 0$ fm$^{-1}$ with the trained RBF network using the s,p,d phase shifts as inputs, and 3000 RBFs.

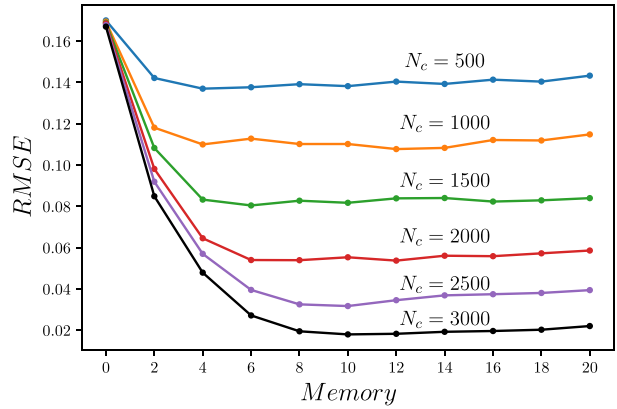

**Fig. 8.** Memory and kernel number dependence of the full RBF network trained with finite potentials. The different colors represent the number of kernels ($N_c$) each having a similiar memory dependence.

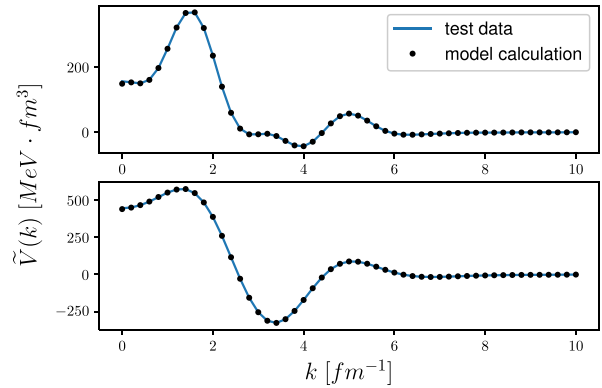

**Fig. 9.** Test of the RBF inversion with two different test potentials in Fourier space, where the blue lines represent the original, while the black dots correspond to the potentials given by the RBF inversion method.

In the next step, we will examine the dependence of the RMSE of the full system shown in Fig. 2 on the memory of the RBF network with a fixed number of s, p, and d input phase shifts and a varying number of centers. The results are shown in Fig. 8, where the different colors correspond to the different number of kernel functions (centers). The same decreasing trend can be seen here as we have observed before with the increasing number of centers; however, now the memory dependence shows a clear minimum in each case, which means it is not necessary to use all of the previous estimated outputs in the following inversion points. In the following, the number of centers will be fixed to $N_c = 3000$, in which case the memory that gives the minimal RMSE will be fixed to $M = 10$. These parameters will be sufficient for all of the follow-up problems as well.

After successfully training the network, we test it by generating two random potentials in coordinate space and calculating their phase shifts, which can be used as inputs for the system. The results are shown in Fig. 9, where the upper and lower panels show the two different potentials in Fourier space (blue lines) with the estimated potentials given by the RBF inversion (black dots). An overall good accuracy has been obtained for both test potentials; however, one still has to be careful due to the error propagation while doing the inverse Fourier transformation, which is necessary to give the potential in coordinate space and calculate the phase shifts by the VPA method. In Sect. 4.2, we will extend our model by using a wider momentum grid and more inversion points, taking into consideration the scattering problem. The training set will also be extended by additional training samples containing potentials with large repulsive (possible infinity) parts near the scattering center.

### 4.2. *Potentials with possible singularities at $r = 0$*

In this section, additional potentials will be generated and used in the training process, which will help the system be able to describe potentials with possible infinities. The model will be extended by considering a larger momentum grid to be able to describe the potentials in coordinate space with a finer resolution. To be consistent with the finite case, we will add the following functional form to the previously generated multisine signals:

$$V_\infty(r) = a\frac{e^{-br^2}}{r^n}, \quad 0 < n \le 1, \quad a, b > 0 \tag{23}$$

where $a$, $b$, $n$ are free parameters controlling the speed of the decay and the magnitude of the repulsive force. When $n = 1$ and $b = 0$ we get the usual Coulomb potential, with $a$ corresponding to the charge of the colliding particles. The Gaussian decay factor is used for practical purposes, as it makes sure that it is consistent with the finite potentials (having the same functional form), and the phase shifts and Fourier transforms could be calculated without numerical issues. The radial Fourier transform of such potentials can be given in the following form:

$$\widetilde{V}_\infty(k) = \frac{4\pi a}{2}b^{\frac{n-3}{2}}\Gamma\left[\frac{3}{2}-\frac{n}{2}\right]{}_1\mathrm{F}_1\left[\frac{3}{2}-\frac{n}{2};\frac{3}{2};-\frac{k^2}{4b}\right] \tag{24}$$

where ${}_1\mathrm{F}_1$ is the confluent hypergeometric function of the first kind defined as:

$${}_1\mathrm{F}_1(a;b;k) = \frac{\Gamma[b]}{\Gamma[b-a]\Gamma[a]}\int_0^1 e^{kt}t^{a-1}(1-t)^{b-a-1}dt. \tag{25}$$

The full potential, which includes the repulsive part, will be given by adding the infinite case to the finite one as follows:

$$V_T(r) = V_0(r) + V_\infty(r), \tag{26}$$

where $V_0(r)$ is the finite, normalized multisine potential given by Eq. (21), while $V_\infty(r)$ is the repulsive part given by Eq. (23). As the radial Fourier transform is a linear operator, the potential in Fourier space can be given by the addition of the Fourier transform of the finite and infinite potentials as:

$$\widetilde{V}_T(k) = \widetilde{V}_0(k) + \widetilde{V}_\infty(k). \tag{27}$$

To train the network, we will generate 10,000 potentials, with half of them containing a repulsive part generated by the previously described method. As now we would like to check the potentials in coordinate space as well, the momentum grid will be extended to $k \in [0, 300]\,\mathrm{fm}^{-1}$. With the generated multisine excitations (including our preliminary knowledge of the system), we could make a "good guess" as to the $\Delta k$ resolution, which is able to capture the full functional form of the radial Fourier transformed potentials, without missing any important detail. From the previous section, when we only used finite potentials with the same multisine form, we knew that $\Delta k = 0.2\,\mathrm{fm}^{-1}$ is a good resolution; however, now $k$ goes up to $300\,\mathrm{fm}^{-1}$, and by using the same small step size throughout the whole grid, the training time would be very long. It is also unnecessary to use a very fine grid above $10{-}20\,\mathrm{fm}^{-1}$, as most of the interesting part happens below that; however, the large-$k$ part also makes an important contribution to the inverse transform, so it has to be included as well. Taking into consideration all the above, the grid is set to be $\Delta k = 0.2$ fm$^{-1}$ between $k = [0, 10]$ fm$^{-1}$, $\Delta k = 0.5$ fm$^{-1}$ between $k = [10, 20]$ fm$^{-1}$, and $\Delta k = 10$ fm$^{-1}$ between $k = [20, 300]$ fm$^{-1}$.

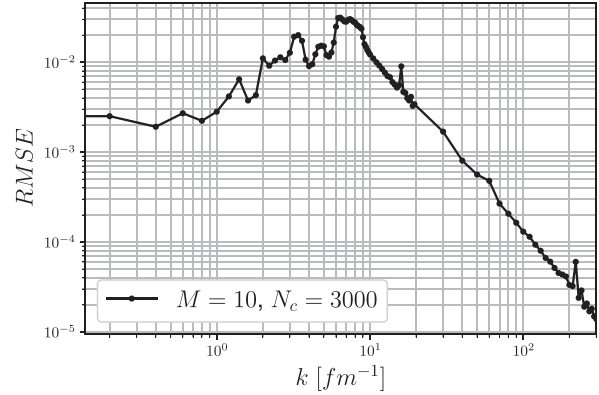

**Fig. 10.** RMSE of the full RBF network using finite and infinite potentials at the different inversion points with $M = 10$ memories and $N_c = 3000$ kernels.

After trying out different network configurations with different complexities, it turns out that including the repulsive parts in the potentials does not make the system more complex, and the same network parameters can be used with the same order of magnitude errors. Considering this, the RBF network is set to have 3 input phase shifts (s,p,d), with an overall $M = 10$ memories, and $N_c = 3000$ basis functions at each inversion point, just as in Sect. 4.1. In Fig. 10 the RMSE measure is shown at each inversion point (momentum grid), showing that the most complex part of the system is between $k \approx [1, 20]$ fm$^{-1}$, where the error is the largest. At small-$k$ the network could learn the system relatively easily; however, as was shown in the previous section, it depends on the number of input phase shifts. At large-$k$ the system becomes very simple, and the errors go down very fast. It has to be noted that it is not necessary to use the same network parameters at each inversion point, e.g. one could easily use simpler networks at large-$k$, giving a much simpler network configuration at the end. Here, it is not necessary to further optimize the network configuration; therefore, in the following problems, we will always use the same complexity at each subnetwork in each inversion point.

The trained network has been tested with two different test potentials, which will be marked by "test data 1" and "test data 2." The results are shown in Fig. 11, where the Fourier transform of the test potentials (blue lines) is compared to the model calculations (black dots). In each case, a good match has been achieved in Fourier space; however, it is necessary to compare the coordinate-space potentials as well. In the upper panel of Fig. 12 the inverse Fourier transformed potential of the function marked by "test data 1" (Fig. 11), and its model estimation are shown, while the lower panel shows the difference between the true and the estimated potentials in coordinate space. The same comparison for the potential marked by "test data 2" can be seen in Fig. 13. The inverse problem is very sensitive at small distances, which could also be seen in both test cases, as in each case the difference in coordinate space is the largest at small distances. The identified model is able to reconstruct the potentials at large $r$ with low errors; however, the small distance parts need some correction, which can be easily achieved by the corrector step described before. To show this, let us apply the RBF corrector to the first test

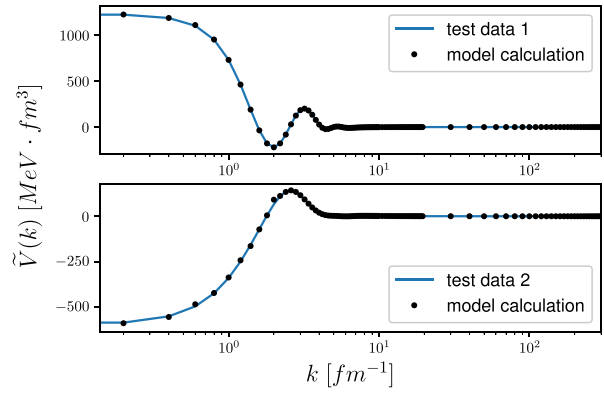

**Fig. 11.** Test of the inversion method for two different test potentials in Fourier space using the extended RBF model with the inclusion of infinite potentials.

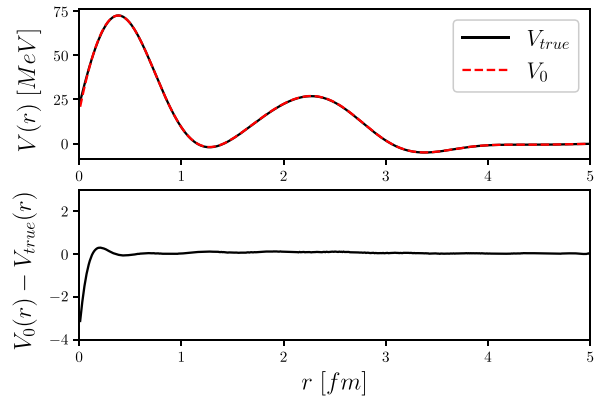

**Fig. 12.** Inverse Fourier transform of the test potential and its model estimation marked by "test data 1" in Fig. 11, where the upper panel shows the potentials in coordinate space, while the lower panel shows the difference between the true and estimated potentials.

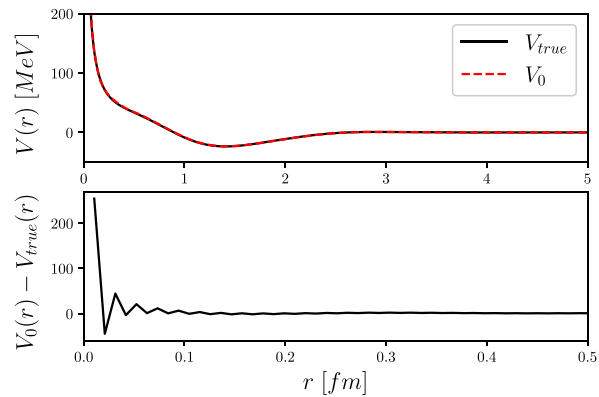

**Fig. 13.** Inverse Fourier transform of the test potential and its model estimation marked by "test data 2" in Fig. 11, where the upper panel shows the potentials in coordinate space, while the lower panel shows the difference between the true and estimated potentials.

potential, "test data 1," with the previously fixed SA parameters. The correction process can be followed in Fig. 14, where the averaged error for the s, p, and d phase shifts at each accepted annealing step is shown. In this case, the correction is stopped when the averaged relative error

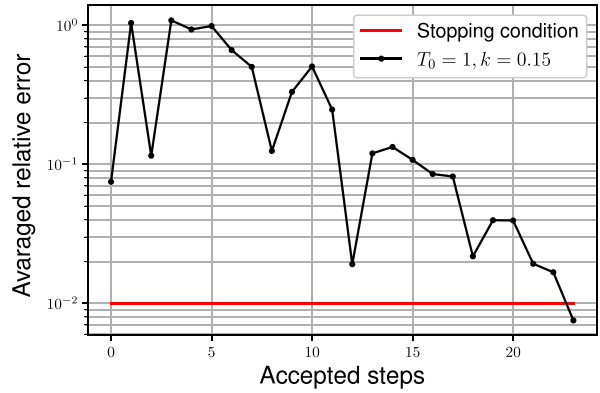

**Fig. 14.** Correction step of the estimated potential corresponding to "test data 1" in Fig. 11, where the black line shows the averaged relative error of the recalculated s,p,d phase shifts at the accepted steps of the SA optimization, while the red line shows the stopping condition corresponding to 1% averaged relative error.

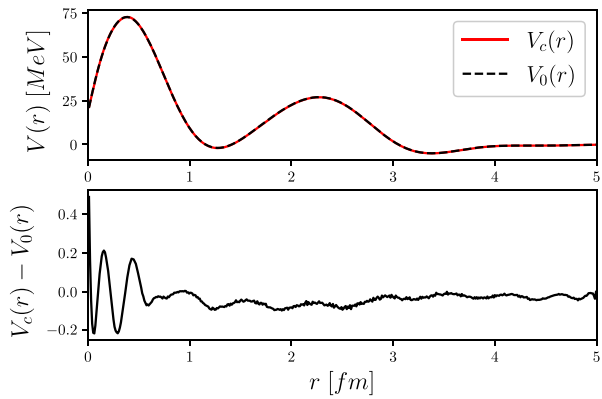

**Fig. 15.** Estimated inversion potential with and without corrections for "test data 1," where the lower panel shows the difference of the two potentials.

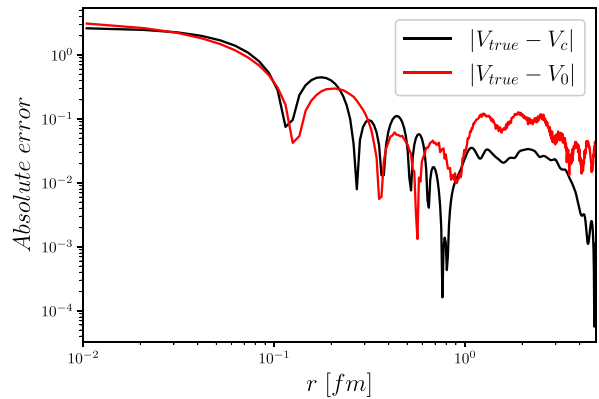

**Fig. 16.** Absolute differences of the estimated potentials (with and without corrections) and the true potential ($V_{\text{true}}(r)$).

reaches 0.01; therefore, the potential obtained by the full neural network inversion (including correction) was able to give the recalculated phase shifts to less than 1% average relative error. The corrected potential ($V_c(r)$) with the original ($V_0(r)$) in coordinate space can be seen in

**Table 1.** Calculated phase shifts from the obtained inversion potentials after the inverse Fourier transform. The "corrected" values refer to the final state RBF optimized results, while the "Inversion" line shows the pure inversion results, without the open-loop RBF control step.

|              | $\delta_0$ | $\delta_1$  | $\delta_2$  |
|--------------|------------|-------------|-------------|
| **True**     | −0.3524    | −0.007538   | −0.001291   |
| **Inversion**| −0.3529    | −0.006744   | −0.001446   |
| **Corrected**| −0.3569    | −0.007608   | −0.001286   |

Fig. 15, while the absolute differences $|V_{\text{true}}(r) - V_c(r)|$ and $|V_{\text{true}}(r) - V_0(r)|$ can be followed on Fig. 16.

The summary of the obtained phase shifts with and without correction is shown in Table 1, where it can be seen that without correction ("Inversion"), the errors in the p- and d-phase shifts are larger but still very close to the true values; however, including the RBF corrector step, the differences in those phase shifts become much smaller. By observing the differences between the uncorrected, corrected, and true potentials in Fig. 16 it is obvious that while the corrector tries to minimize the difference between the calculated and measured phase shifts, the corrected potential will still not be a perfect copy of the true potential. This is a straight consequence of the fact that the phase shifts are not that sensitive to variations at small distances; thus, the corrector will prefer changes at intervals where it is easier to achieve better results faster. According to this, the most sensitive part is the interval after 1 fm, so the corrector will prefer to make changes at those parts. Due to the very sensitive nature of the inverse problem, it is easy to minimize the phase shift differences by just adjusting the "almost good" potential with small corrections. Without additional knowledge of the true potential, it is very hard to make better corrections; however, it has to be noted that the overall relative errors are still very small, and after corrections, they became even smaller. For the uncorrected potential, the average relative error between the estimated and true potential is around 7%, while the same error for the corrected potential is around 1%, which is very good considering that in a real measurement we always have errors, which in many cases are larger than a few percent. On a side note, it is worth mentioning that if we have more information about the scattering system, e.g. bound state energies, values of different moments of the potentials in coordinate space, etc., it would be possible to construct an optimizer that takes into account this extra information as constraints, therefore making it possible for the algorithm to, e.g. focus on the small distance parts more than the large distance parts.

In the next section, the RBF inversion method will be applied to neutron–α scattering, where the scattering potential will be determined in both Fourier and coordinate space.

## 5. Application: estimating the central *n−α* interaction potential with the RBF method

In this section, we will apply the neural network inversion to neutron–α scattering at $T_{\text{lab}} = 10$ MeV incident energy. The scattering problem is very similar to the previously shown problems; therefore, the momentum resolution and the type of training samples could still be used here as well. However, as the masses and spins of the particles are different, new training and test samples have to be generated. Additionally, we have to consider the spins of the scattering particles, which has not been done before. In this case, the spin 1/2 neutrons are scattering on the spinless α particles; therefore, we can split the potential into a central and a spin-orbit term

as [65]:

$$V(r) = V_{\text{central}}(r) + a_l^\pm V_{\text{so}}(r), \tag{28}$$

where $V_{\text{central}}(r)$ is the pure central component, and $V_{\text{so}}(r)$ is the spin-orbit term of the scattering potential, while the coefficient of the spin-orbit term is:

$$a_l^\pm = \frac{1}{\hbar^2}\langle \mathbf{s}\mathbf{L}\rangle, \tag{29}$$

which, depending on the spin configuration, could be $l$ for $j = l + 1/2$ or $-(l + 1)$ for $j = l - 1/2$.

In simple scattering scenarios at low energies, the simplest thing one could do is approximate the scattering process with the interaction of spinless particles and calculate an effective radial potential, which could describe the scattering phase shifts. On the other hand, it is also possible to identify both the central and the spin-orbit terms by generating training samples that contain both terms. The latter method would be necessary if one wanted to describe the system in more detail; however, in this case, we would like to describe only the central part of the potential approximately by not forgetting the spin dependence as well. This could be achieved by the so-called LHF approximation [66], which takes into account the spin–orbit interaction by assuming a weak spin–orbit coupling. The LHF approach uses the assumption that the scattering of such systems can be approximated by the Distorted Wave Born Approximation (DWBA) [67], and the scattering phase shifts of the up and down components can be expanded by the spin–orbit coupling term $a_l^\pm$ as:

$$\delta_l^\pm = \delta_l^c + a_l^\pm C_l^{(1)}(k) + \left(a_l^\pm\right)^2 C_l^{(2)}(k) + \mathcal{O}\left(\left(a_l^\pm\right)^3\right). \tag{30}$$

From $\delta_l^\pm$ we can define the following quasi-independent phase shifts:

$$\tilde{\delta}_l = \frac{1}{2l + 1}\Big[l(l + 1)\delta_l^+ + l\delta_l^-\Big], \tag{31}$$

and

$$\hat{\delta}_l = \frac{1}{2l + 1}\Big[l\delta_l^+ + (l + 1)\delta_l^-\Big], \tag{32}$$

where $\delta_l^+$ and $\delta_l^-$ can be expanded in powers of the angular momentum $l$ as shown in Eqs. (29,30). Keeping only the first-order terms, the quasi-independent $\tilde{\delta}_l$ and $\hat{\delta}_l$ phase shifts can be expressed as $\tilde{\delta}_l \sim \delta_l^{(0)}$ and $\hat{\delta}_l \sim \delta_l^{(0)} - C^{(1)}$, respectively, where $\delta_l^{(0)}$ corresponds to the pure central part of the potential $V_{\text{central}}(r)$, while $C_l^{(1)}$ corresponds to half of the spin-orbit term $1/2 V_{\text{so}}$. Starting from these assumptions, our task is to determine the quasi-independent $\tilde{\delta}_l$ and $\hat{\delta}_l$ phase shifts from the measured $\delta_l^\pm$ phase shifts and then estimate $V_{\text{central}}(r)$ from $\tilde{\delta}_l$. To estimate the spin-orbit term, we need to calculate the $\hat{V}_l(r)$ potential corresponding to the $\hat{\delta}_l$ phase shifts, then by using the first-order approximation and the previously estimated $V_{\text{central}}(r)$ central potential, the spin-orbit term can be estimated as $V_{\text{so}}(r) = 2V_{\text{central}}(r) - 2\hat{V}(r)$. Here, we will only estimate the central part; however, the spin-orbit term could be approximated by the same method using the same RBF network, and there is no need for new training samples or another training step.

After the short theoretical background, let us proceed with the training of the RBF network, which will be used to give a first estimation of the central potentials. The system now consists of two particles with different masses: $m_n = 940$ MeV and $m_\alpha = 3720$ MeV, and the scattering energy has been set to $T_{\text{lab}} = 10$ MeV. The momentum grid is set as before: $\Delta k = 0.2$ fm$^{-1}$ between $k = [0, 10]$ fm$^{-1}$, $\Delta k = 0.5$ fm$^{-1}$ between $k = [10, 20]$ fm$^{-1}$, and $\Delta k = 10$ fm$^{-1}$ between $k = [20, 300] \, \text{fm}^{-1}$. To train the network, 5000 finite multisine, and 5000 short-distance repulsive multisine potentials were generated with the same parameters as were described in Sect. 4. The RMSE of the trained network for 2000 test potentials is shown in Fig. 17, giving the same order of magnitude errors as that which we observed before, suggesting that the proposed network structure could describe the underlying system very well.

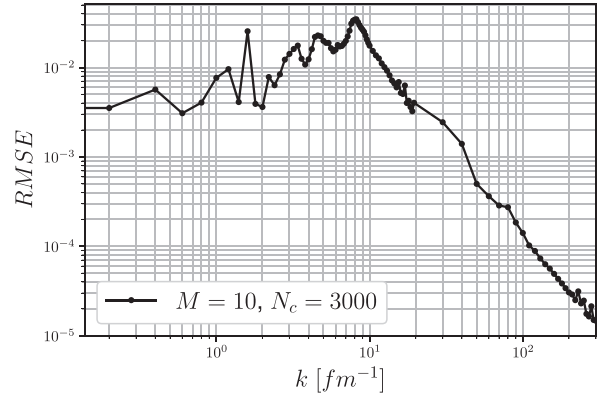

**Fig. 17.** RMSE of the trained RBF network with $M = 10$ memories and $N_c = 3000$ kernel functions at each momentum grid point, applied to $n-\alpha$ scattering using 2000 test samples.

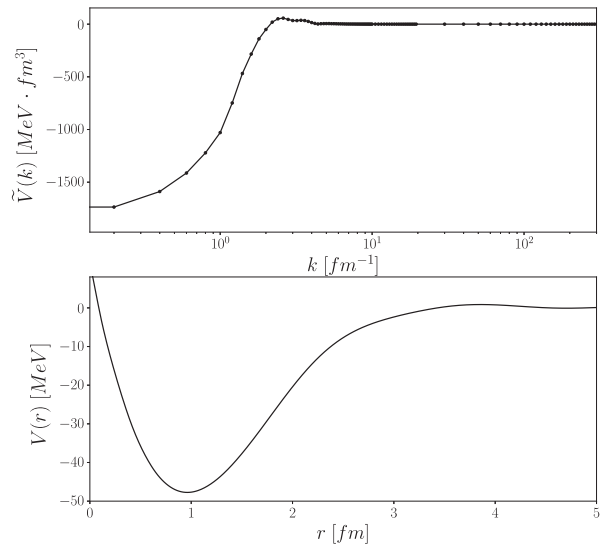

**Fig. 18.** Estimated central potentials in Fourier space (upper panel) and in coordinate space (lower panel) obtained by the RBF inversion method without final state correction using the measured s, p, d phase shifts.

The trained network can now be applied to the measured data collected in Refs. [68,69], where the $\tilde{\delta}_l$ phase shifts have been determined to be $\tilde{\delta}_0 = 1.87$, $\tilde{\delta}_1 = 1.62$, $\tilde{\delta}_2 = 0.038$. Using these values, the central potential can be estimated. The results of the inversion in Fourier space and its inverse Fourier transformed coordinate-space potential can be seen in Fig. 18. After obtaining the uncorrected inversion potential from the RBF inversion block, it has to be pushed through the RBF corrector block to make the necessary corrections for the recalculated phase shifts. The correction process can be seen in Fig. 19, where the same parameters have been used as before. The results for the full inversion are summarized in Fig. 20, and in Table 2, where the latter shows the corrected inversion potential in coordinate space, while the former shows the recalculated phase shifts of the corrected inversion potential compared to the original phase shifts.

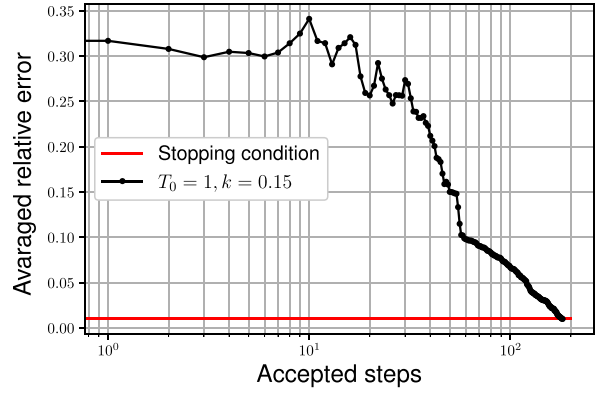

**Fig. 19.** Correction of the estimated n–α central potential by the RBF corrector using SA optimization. The stopping condition corresponds to 1% averaged relative error for the s, p, d phase shifts.

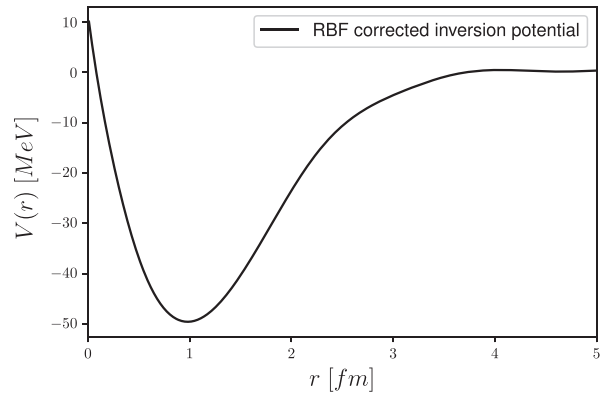

**Fig. 20.** The estimated $V_{\text{central}}(r)$ central potential for n–α scattering at $T_{lab} = 10$ MeV energy using the RBF neural network inversion model with the SA corrector in the final step.

**Table 2.** Comparison of the measured n–α phase shifts with the results from the RBF corrected neural network inversion method.

|  | $\tilde{\delta}_0$ | $\tilde{\delta}_1$ | $\tilde{\delta}_2$ |
|---|---|---|---|
| **Meas.** | 1.87 | 1.62 | 0.038 |
| **Inv. + corr.** | 1.921 | 1.618 | 0.0379 |

The results show the expected behavior of the nuclear potential, as it has a very short-range repulsive and a medium-range attractive part with a minimum around $r \approx 1$ fm. The potential dies out around $4-5$ fm, which is also expected. The preliminary knowledge of the system greatly simplified the overall identification process; however, when we do not have a detailed preliminary knowledge of the system, it is still possible to construct a well-suited network that will be able to grasp the underlying system. In that case, however, the overall training procedure could be much more complicated, so the identified system could be more complicated as well, which, on the other hand, will be able to describe a wider range of scattering systems.

During the identification process, only decaying multisines with and without short-range repulsive parts were used, which could raise the question of the generalization capabilities to other types of functional forms as well. In general, when we want to identify a linear or nonlinear system, we want to make sure that the system is excited at all of the important frequencies in a sensible range to be able to measure its response and make assessments on the behavior of the system. The identification process in the nonlinear case is much harder because the superposition principle does not hold anymore, and the system could shift energy from one frequency to another, making the overall behavior of the system much more complicated than in the linear case. Random phase multisines are specifically designed to excite nonlinear systems in a more controlled manner than, e.g. simple noise excitations, while still maintaining a specific randomness, which is necessary for a good generalization at the end. The training of neural networks is similar to other system identification methods, and the training samples have to be able to excite the system in the operating range for which we would like to make estimations. In our case, the expected signals have to be continuous, decaying functions, so any functional form could be used for training, e.g. polynomials, sums of Gaussians, sums of sines and cosines, etc. Multisines were a straightforward choice to generate good training samples with the desired properties; however, this should not mean that the trained neural network should only be able to give responses that are multisine signals. One of the good properties of neural networks is their generalization capabilities, which here would mean that the trained system will be able to interpolate between the training functions, or, in other words, interpolate between the different frequency responses. The system should be able to give approximately correct answers to functional forms other than multisines, and the response of the system should not be dependent on the specific functional form of the training samples. Through the operating range, however, the system still depends on the training samples; e.g. it would have a hard time generalizing to noncontinuous potentials, which were not part of the training process, and its frequency response would be outside of the scope of the currently trained network. The network would still try to generalize to these inputs; however, it would mean the system should extrapolate to intervals where we could not guarantee a good generalization. By using the trained system in the designed operating range, however, it would mean that it is able to interpolate enough so that not just random phase multisines but any sensible potential can be approximated with good accuracy.

## 6.  Conclusions

In general, solving the inverse scattering problem in quantum mechanics is a tedious task due to its high sensitivity to small perturbations, nonlinearities, and/or nonuniqueness. In many cases, the problems could be constrained so that usual numerical methods could be applied for solving the integral equations arising in the description of the inverse dynamical systems; however, due to the high sensitivity and the finite number of measurements, in practice it is impossible to fully describe the problems.

In this paper, we consider the inverse scattering problem at fixed energies, where a finite number of measured phase shifts are used to estimate the interaction potentials. To determine the scattering potentials in fixed-energy experiments, an RBF-type neural network structure is constructed, which is then trained by well-defined multisine potentials to identify the underlying nonlinear system. The estimated potential is corrected by an RBF-type corrector using SA as an optimizer to correct for the differences in the recalculated and measured phase shifts, assum-ing that only small perturbations are needed to achieve smaller errors. The method was able to produce less than 1% averaged relative errors for the recalculated s, p, and d phase shifts for the test samples.

To test the inversion scheme on real-life data, we used the measured s, p, and d phase shifts for neutron–$\alpha$ scattering at $T_{\text{lab}} = 10$ MeV incident energy, where we took into consideration the spins of the scattering particles approximately by the LHF approximation. By using the measured phase shifts, the central part of the interaction potential could be expressed, and the RBF network could be used in the same way as it would be for spinless particles. By using the RBF inversion and the corrector, a physically sensible potential has been obtained with less than 1% averaged relative error for the recalculated and measured s, p, d phase shifts.

The method is flexible enough to be able to describe different systems with different masses, spins, energies, etc. It is also possible to construct a network that could describe the fixed-energy inversion at many energies by introducing the energy as an input to the full system. In this model, RBF-type networks have been used; however, any other type of neural network could be used as well. The neural network inversion model could possibly be used to describe the inverse problem of inelastic scattering as well, where one should also consider the imaginary part of the interaction potentials. Due to the flexibility of the proposed method, there are many possible future applications, which could help us understand more complex problems as well.

### Acknowledgements

This work was supported by the National Research Foundation of Korea (NRF) grant (No. 2018R1A5A1025563) funded by the Korea government (MSIT). The author was supported by the Hungarian OTKA fund K138277.